# Predicting Safe Regions within Lava Flows over Topography

**Jack M. Saville[1], Edward M. Hinton[2], and Herbert E. Huppert[3]**

[1] Department of Earth Sciences, University of Cambridge, CB2 3EQ, UK.

[2] School of Mathematics and Statistics, The University of Melbourne, Parkville, VIC 3010, Australia.

[3] Institute of Theoretical Geophysics, King's College, University of Cambridge, CB2 1ST, UK.

Corresponding author: Jack M. Saville (js2509@cam.ac.uk).

**Key Points:**

- Safe regions, into which lava does not flow, form downslope of sufficiently elevated features. Fast, deep flow is diverted to their margins.

- Lava ponds in deep depressions. Flow is delayed by the time taken to fill the depression. The region downstream stays lava-free for longer.

- Our model is compared to the lava field at Marcath Volcano and replicates an observed lava-free region downslope of a topographic high

**Abstract**

We present a shallow, isothermal, Newtonian model for the transient interaction of lava flows with topography. Numerical integrations and simple mathematical approximations are deployed to quantify how topography controls lava thicknesses and flow speeds. Considering idealised topographic features, we show that modest depressions thicken and accelerate the flow - even far downstream - whilst mounds have the opposite effect. However, deep ponds of lava form in depressions of sufficient amplitude, which introduces a long timescale for lava to fill the depression and hence the accelerated downstream flow may never be attained. Relatively large mounds completely divert the lava, providing protected lava-free regions for homes and infrastructure. There can, however, be hazardous, deep, fast flow around the edges of the mound, owing to diversion. Additionally, we show that our model accurately predicts the lava-free region that has been observed in the eruption 35 kyr ago at Marcath Volcano, Nevada.

**Plain Language Summary**

Lava flows in volcanic regions cause huge amounts of economic damage. We develop a model to predict the depth and speed of lava flows over uneven surfaces, aiming to locate safe regions in which homes and infrastructure are protected from the flow. We show that mounds both thin and slow the flow of lava over them, with lava-free safe regions forming downslope of larger mounds. Depressions have the opposite effect, thickening and accelerating the lava flow. Lava ponds within deep depressions, which can delay the progress of the flow, giving occupants of the area longer to evacuate, and potentially protecting them for the entire duration of an eruption. Additionally, we apply our model to a real scenario, showing that it reproduces the lava-free region that has been observed at Marcath Volcano, Nevada.

# 1 Introduction

Accurate simulations of lava emplacement are vital for lava flow forecasting, assessing volcanic hazard, and designing barriers or other mitigation methods. Lava often flows over topography as rough brecciated basaltic flows known as a'ā (Macdonald, 1972; Kilburn & Luongo, 1993). These are observed at Etna (e.g. Guest et al., 1974), Hawaii (e.g. Lipman & Banks, 1984) and Iceland (e.g. Rossi, 1997) among other volcanically active localities. Such flows can travel large distances and have the potential to cause huge economic damage if they flow into populated areas (e.g. Chirico et al., 2009). They also pose a threat to life, with approximately one thousand fatalities pertaining directly to lava flows between 1600 and 2010, (Auker et al., 2013). Accurately and rapidly predicting the behavior of lava flows in the early stages of an effusive eruption is critical for the proactive evacuation of people and possessions from the flow's path. Furthermore, the forecasted lava-free regions behind sufficiently large features in the flow's path (e.g. Younger et al., 2019) can act as safe regions during a lava flow, offering protection for homes and infrastructure (Jenkins et al., 2017). As suggested by Moore (1982) and Scifoni et al. (2010), these findings inform the safest positioning of homes and infrastructure with respect to topographic features.

The rheological complexity of lava - which cools and crystallises as it travels - makes simulating lava flows challenging (Lyman & Kerr, 2006) because the accompanying increase in viscosity at flow margins can modify the shape of a flow (Stasiuk et al., 1993; Kerr et al., 2006). The dominant control on the shape of a lava flow, however, is the underlying topography. Existing simulations of lava flows vary in their physical completeness and computational expense. Some are tailor-made for lava flows while others make use of existing computational fluid dynamics software (Cordonnier et al., 2016; Dieterich et al., 2017).

The simplest useful lava flow model is the steepest-descent method, where lava is assumed to flow indefinitely over topography along paths of steepest descent. This methods affords very short computational times, and stochastic modifications to topography between runs have been used to produce probabilistic flow forecasts (Favalli et al., 2005), modelling the likely flow direction but not the temporal evolution or thickness of the flow. Though useful for volcanic hazard mapping, this simplistic method only considers the component of gravity acting downslope: the component acting into the slope, which acts to smooth out variations in flow thickness, is neglected. Hence the steepest-descent method struggles to predict the closure of lava-free regions behind many obstacles. Furthermore, the method is valid only in the limit of thin flows, with gradients in flow thickness much smaller than topographic slopes. This ensures the smoothing effect is negligible relative to topographic forcing. This is rarely true at the steep margins of a'ā flows. When lava flows into topographic depressions, deep ponds of lava have been observed to form (Pedersen et al., 2017), whose behavior cannot be satisfactorily captured by the steepest-descent method.

The most complete lava flow models simulate three-dimensional viscous flow with spatially and temporally varying rheology, accounting for complex aspects of lava physics such as crystallization and degassing (e.g. Hidaka et al., 2005; Herault et al., 2011). Such models can reproduce intricate features of real lava flows, but their long computation times render them unable to simulate flows quickly, making them less useful for active eruption forecasting. Such models provide little insight into the effect of topography or other parameters on lava flows, because their inputs cannot be tweaked and quickly rerun multiple times.

To reduce computation time, aspects of lava physics must be neglected. Following Lister (1992), many models use a depth-averaged approach for shallow flows, which reduces the computational domain to two dimensions (e.g. Costa & Macedonio, 2005; Kelfoun & Vargas, 2016). Other models use a cellular automata approach, where lava spreads between neighboring cells in a two-dimensional grid depending on their relative elevations, lava thicknesses and other parameters (e.g. Crisci et al., 1986; Young & Wadge 1990; Vicari et al., 2007; Del Negro et al., 2008). Many two-dimensional approaches include a thermal model and temperature dependent rheology but cannot fully capture cooling and crust formation. Further dimensional reduction allows modelling of one-dimensional flow along a channel (e.g. Harris & Rowland, 2001). This reduces computation time and can give valuable insight into thermal and rheological evolution of lava flows but cannot account for the effect of topography.

Notably lacking from the literature is a simplified analysis of the interaction between a'ā flows and both mathematically idealized and real-world topographic features. In this work, a two-dimensional model with rapid computation time is developed, in which parameters can be varied between runs to investigate the interplay of topography, flux and viscosity on lava flow dynamics. The shallow-flow approximations of lubrication theory are applied to an isoviscous, isothermal Newtonian fluid. This reflects the approach successfully taken by Huppert (1982), Huppert et al. (1982), and Lister (1992). Our model provides quick-to-run, accurate, two-dimensional simulations of both the temporal evolution and steady-state behavior of a'ā flows, and is able to simulate ponding of lava in depressions, unlike the steepest-descent method. Inertia and surface tension are neglected, since they are unimportant away from the vent (Walker & Mullins, 1981; Lev et al., 2012). The steady version of this approach was taken by Hinton et al. (2019; 2020) for flow over idealised topographic features consisting of isolated Gaussian mounds and depressions on inclined planes. In section 2, we introduce a time-dependent model for lava flows. By analysing the temporal evolution of the lava, we identify which topographies act to retard or accelerate the flow, a feature that was not captured by the previous steady research. In sections 3 and 4, we investigate the flow over topographic features which combine mounds and depressions. By considering interactions with idealised topography, we derive some general principles describing the rate of downslope migration of the flow front, the steady flow depth behind the front, and the locations of any lava-free regions that form. We then show that these principles can be applied to real topography by comparing our model to the distribution of ancient lava flows emplaced in an eruption 35 kyr ago at Marcath Volcano, Nevada, observed by Younger et al. (2019). An animation of the modelled flow at Marcath is provided in this paper's supplementary information.

## 2 Model Formulation

To quantify how lava flows are diverted, focused and slowed by different topographic features, we first develop a mathematical model. Observed lava flows are much shallower than both their width and the length scale over which natural topography has large variations, inspiring a shallow-flow approximation to the Navier-Stokes equations, termed lubrication theory (Batchelor, 1967). We model lava as an isothermal Newtonian fluid with constant density and viscosity. A steady flux is introduced from a line or point source at a topographic high and the flow depth is analysed during the subsequent gravity-driven free-surface flow. We introduce the notation listed in Table 1.

| Variable | Quantity | Units |
|---|---|---|
| $(x, y, z)$ | Spatial coordinates | $m$ |
| $t$ | Time | $s$ |
| $h(x, y, t)$ | Vertical thickness of lava | $m$ |
| $E(x, y)$ | Vertical height of underlying topography | $m$ |
| $\mu = 5 \times 10^6$ | Dynamic viscosity of lava | $Pa\ s$ |
| $\rho = 2650$ | Density of lava | $kg\ m^{-3}$ |
| $g = 9.81$ | Local gravitational acceleration | $m\ s^{-2}$ |
| $h_\infty$ | Lava depth at upstream line source | $m$ |
| $\beta = 10°$ | Angle of inclined plane to horizontal | dimensionless |
| $m(x, y)$ | Profile of feature on inclined plane | $m$ |
| $q$ | Volumetric lava effusion rate | $m^3 s^{-1}$ |

**Table 1:** A summary of the mathematical notation used in this paper. Numerical values reflect estimates taken from Younger et al. (2019). The viscosity estimate falls within the range suggested for a'ā lavas by Belousov and Belousova (2018).

For shallow viscous flow, the velocity profile within the fluid is parabolic with depth (Batchelor, 1967). The velocity is integrated over the fluid's thickness to obtain the volume flux, which is combined with volume continuity to obtain the governing partial differential equation (1) for the flow thickness, $h(x, y, t)$ over topography with vertical height $E(x, y)$. For mathematical details see Hinton et al. (2019).

$$\frac{\partial h}{\partial t} = \frac{\rho g}{3\mu} \nabla \cdot [h^3 \nabla (E + h)]$$

(1)

The first term on the right-hand side of (1), $\nabla \cdot [h^3 \nabla E]$, represents the effect of gradients in underlying topography, describing downslope movement of fluid. The second term, $\nabla \cdot [h^3 \nabla h]$, represents the change in fluid depth pertaining to the component of gravity perpendicular to the slope, which smooths out gradients in flow thickness.

We are also interested in the behaviour of the lava flow at long times after the transient flow front has passed the topographic feature. It has been shown analytically by Lister (1992) that flows of this type with input of fluid at a constant rate reach a steady depth upstream of the flow front, and observations suggest this is also the case for real lava flows (Macdonald, 1972; Kilburn & Luongo, 1993). This motivates analysing the steady problem, in which the time derivative in (1) vanishes to give the steady governing equation

$$\nabla \cdot [h^3 \nabla(E + h)] = 0. \tag{2}$$

Equation (1) or (2) are solved numerically for a given topographic surface $E(x, y)$. Details of the numerical methods used are given in the appendix.

In the case that gradients in flow thickness are much shallower than topographic gradients ($|\nabla E| \gg |\nabla h|$), the second term in (1) is negligible and the resulting time-dependent governing equation (3) is purely forced by topography. Similar to the steepest-descent method discussed above (Favalli et al., 2005), the term describing smoothing of gradients in flow thickness by the component of gravity into the slope is neglected, meaning this approximation predicts lava to flow along paths of steepest descent, with

$$\frac{\partial h}{\partial t} = \frac{\rho g}{3\mu} \nabla \cdot [h^3 \nabla E]. \tag{3}$$

Unlike previous work using the steepest-descent method, equation (3) describes the temporal evolution of flow thickness in addition to the flow path. Using the method of characteristics (Evans, 2010), equation (3) is reduced to a system of ordinary differential equations describing a 'characteristic' lava trajectory (4a) and flow thickness along that trajectory (4b).

$$\frac{d}{dt}(x, y) = -\frac{\rho g}{\mu} h^2 \nabla E \qquad \text{and} \qquad \frac{dh}{dt} = \frac{\rho g}{3\mu} h^3 \nabla^2 E. \tag{4a, 4b}$$

For given topography $E(x, y)$, this system is numerically integrated forwards in time with the upstream boundary condition fixing the flow depth at a set of points at the source. This furnishes a set of curves, each one representing the flow path from a different point at the source. The flow front at any given time is then obtained from this set of curves.

Equations (4a,4b) are valid provided that the topographic surface does not contain uphill regions (and so is fully coated by fluid at steady state). Uphill flow is not possible in the absence of inertia. Instead,the fluid either forms a deep pond to overtop the uphill area (such deepening is not allowed in the characteristic approximation) or is entirely diverted around the uphill area. These two modes of model breakdown are indicated either by convergence of characteristic curves to a point, indicating pond formation in a depression, or by the development of regions that are entirely inaccessible to characteristic curves, indicating formation of a lava-free region in the wake of a mound. The steady version of these phenomena have been studied analytically for the case of flow over a single Gaussian mound or depression by Hinton et al. (2019; 2020).

As an example, a topographic feature with profile $m(x, y)$ is placed on a plane inclined at angle $\beta$ to the horizontal such that the topographic elevation profile has the form $E(x, y) = -x \tan \beta + m(x, y)$. Here, we use $\beta = 10°$, which is applicable to many natural slopes, including those reported at Mauna Kea by Mark and Moore (1987) and at Marcath Volcano by Younger et al. (2019). However, it should be noted that our results apply to any slope and scale of feature. The steady thickness of the lava flow depends qualtitatively only on the ratio of the feature's gradient to the slope gradient, which is discussed in more detail in section 3.

Presented in figure 1 is the investigation of an isolated Gaussian mound and depression on inclined planes wherethe topographic profile has the form

$$E(x, y) = -x \tan \beta \pm D \exp\left[-\frac{x^2 + y^2}{L^2}\right], \qquad (5)$$

and the feature has radius $L = 100$ m and height $D = 10$ m. These values are selected to probe the suitablilty of the model for natural topographic features. Figures 1a and 1b show topographic profiles for the mound and depression respectively, while the characteristic solutions to (4) for these topographies with a laterally infinte upstream line source where lava has depth $h_\infty = 5$ m are presented in figures 1c and 1d. The corresponding steady numerical solutions to (2) are shown in figures 1e and 1f.

By inspection of (4) and figures 1c and 1d, we find some general results. Flow over the centre of a mound, where $\nabla^2 E < 0$, is thinner and slower, represented by sparse characteristics and slow progress of the flow front over the mound, while fast, deep flow is diverted to its margins. Flow over a depression, where $\nabla^2 E > 0$, is thicker and faster, represented by dense characteristics and rapid progress of the flow front. We also notice that this model predicts flow to be along curves of steepest descent on the surface, because $d(x, y)/dt \propto -\nabla E$. This makes physical sense, because a vanishingly thin flow is forced entirely by the underlying topography and flows downhill, perpendicular to topographic contours. We initially consider only topography and lava sources that are symmetrical about $y = 0$. We restrict our analysis to the $y > 0$ half-domain, and can reconstruct the lava thickness over the whole domain by symmetry.

It should be noted that the viscosity and density influence only the rate of flow in equations (1), (3) and (4), not the direction, which depends solely on the topography. Hence, although real lavas are not isoviscous, our model is able to accurately capture the qualitative flow over topography (e.g. section 4). Indeed, even non-Newtonian fluids will generally follow the direction of steepest descent and the steady flow thickness will be qualitatively similar to the Newtonian case, as has been shown for a yield-stress fluid (Hinton & Hogg, 2022). In addition, since viscosity and density only influence the timescale of the flow, they effect the steady state that arises at very late times only via the flow thickness at the source, which is a function of $\lambda = q\mu/\rho$ (Lister, 1992). Even in transient lava flows, varying $\mu$ but holding $\lambda$ fixed will only change the timescale of the flow, not the qualitative features of the flow thickness (for further details, see section 4).

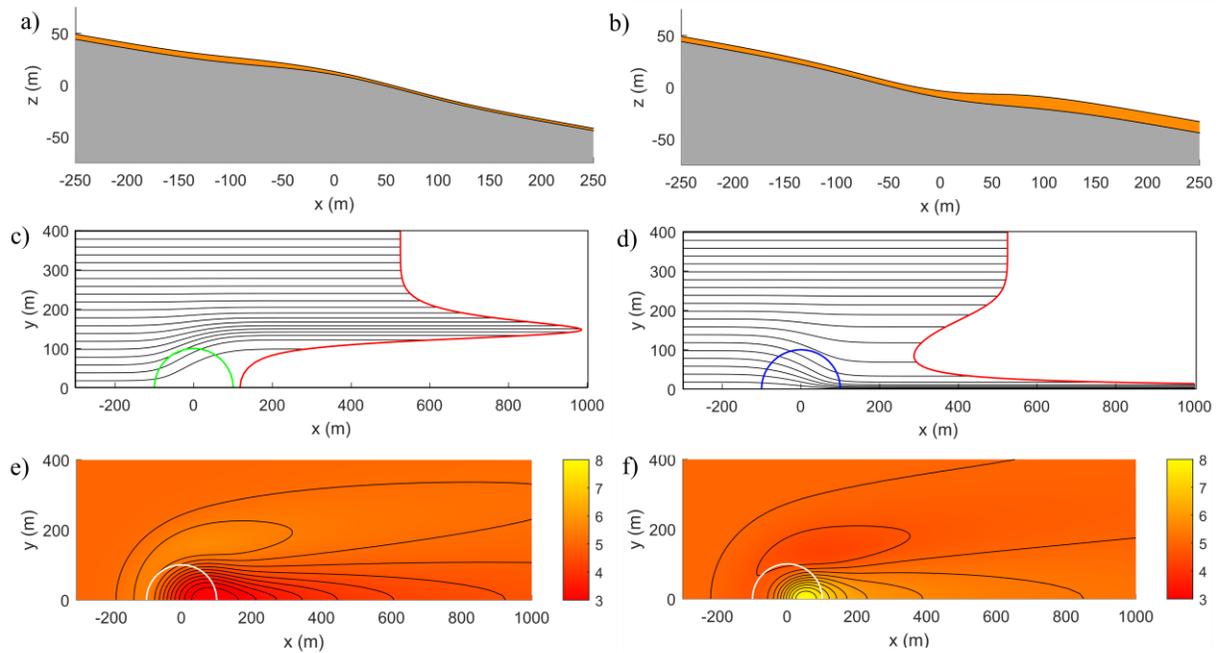

**Figure 1:** Flow over **(a)** a Gaussian mound and **(b)** a Gaussian depression, both with radius $L = 100$ m and height $D = 10$ m on planes inclined at $\beta = 10°$ to the horizontal, inundated from a line source at $x = -300$ m, where the lava has depth $h_\infty = 5$ m. Lava depth is indicated by the orange layer. The characteristic solutions to (4) for **(c)** the mound and **(d)** the depression show the area covered by the flow at $t = 2.5$ hours (only $y > 0$ is shown because the flow is symmetric about $y = 0$). The shape of the transient flow front is indicated by the red curve joining the downstream ends of characteristics, and the extent of each feature is indicated by a semicircle. The numerical solutions to (2) are presented for **(e)** the mound and **(f)** the depression. The colour represents the steady flow thickness in metres.

## 3 Idealised Topographic Features

With a model to describe lava flow over general topography in hand, we next present some general principles based on the study of idealised topographic features that contain both mounds and depressions. Previous theoretical work by Hinton et al. (2019; 2020) found that steady flow is diverted around a Gaussian mound (with sufficiently steep mounds producing lava-free regions), and is focused towards a Gaussian depression (with sufficiently steep depressions producing deep ponds of lava; see figure 1). Here, we consider more complex topographic features to provide deeper insight into the effects of topography on steady flows. We also analyse the transient evolution of flow over topography, finding that larger features may not always be the optimal defence. In the following subsections, we investigate the effect of a mound and a depression in series (section 3.1) and a mound with a depression at its centre (section 3.2). Finally, in section 4, the ideas are deployed to model real flow over topography from the eruption at Marcath volcano, 35 kry ago.

### 3.1. Mound and Depression in Series

As shown in figure 1, relatively small mounds divert the flow and relatively small depressions focus the flow. This complementary behaviour motivates the study of a mound and a depression in series. We consider the shallow steady flow of lava over a Gaussian mound followed by an identical but inverted Gaussian depression on an inclined plane, separated by a downslope distance $x_{offset}$. For such flow over an arbitrary feature $m(x, y)$ on a plane inclined at $\beta$ to the horizontal, the steady version of equation (3) becomes

$$0 = \nabla \cdot \{h^3 \nabla [-x \tan \beta + m(x, y)]\}. \tag{6}$$

We notice that (6) is invariant under the transformation $x \rightarrow -x$, $m \rightarrow -m$, implying that the steady flow over an inverted feature $-m(x, y)$ is found by reflecting in the $y$-axis the steady flow over a feature $m(x, y)$. Hence, at steady state, an inverted feature cancels the effect of a non-inverted feature such that if a feature is followed by its negative, the flow depth is returned to its unperturbed value downstream of both features. This result generalizes to non-Newtonian rheology provided that the constitutive law does not depend explicity on time. Flow depth is represented by the density of characteristics in figure 2, where a Gaussian mound in series with a Gaussian depression on an inclined plane is investigated. The topographic elevation profile has the form

$$E(x, y) = -x \tan \beta + D \exp\left[-\frac{x^2 + y^2}{L^2}\right] - D \exp\left[-\frac{(x - x_{offset})^2 + y^2}{L^2}\right], \tag{7}$$

with height $D = 6$ m, radius $L = 100$ m, feature spacing $x_{offset} = 300$ m, source depth $h_\infty = 5$ m and underlying slope $\beta = 10°$. The density of characteristics returns to its original value downstream of both features, irrespective of their ordering, as indicated in figures 2a and 2b.

Although the flow depth is returned to its unperturbed value downstream of the second feature, the same cannot be said for the shape of the transient flow front, which depends on the ordering of the features and their separation. Flow is fastest where it is thickened, which is represented by the concentration of characteristics. The flow front becomes perturbed when flow depth changes along characteristics. Delay of the flow front along the centreline, $y = 0$, occurs in flow over a mound because a region of thinned flow is created in the wake of the mound, which slows the front. A depression further downstream returns the flow to its unperturbed depth and velocity, but the flow along the centreline has been permanently delayed, owing to the thinned flow between the features as shown in figure 2a). The converse occurs when a mound follows a depression, as shown in figure 2b.

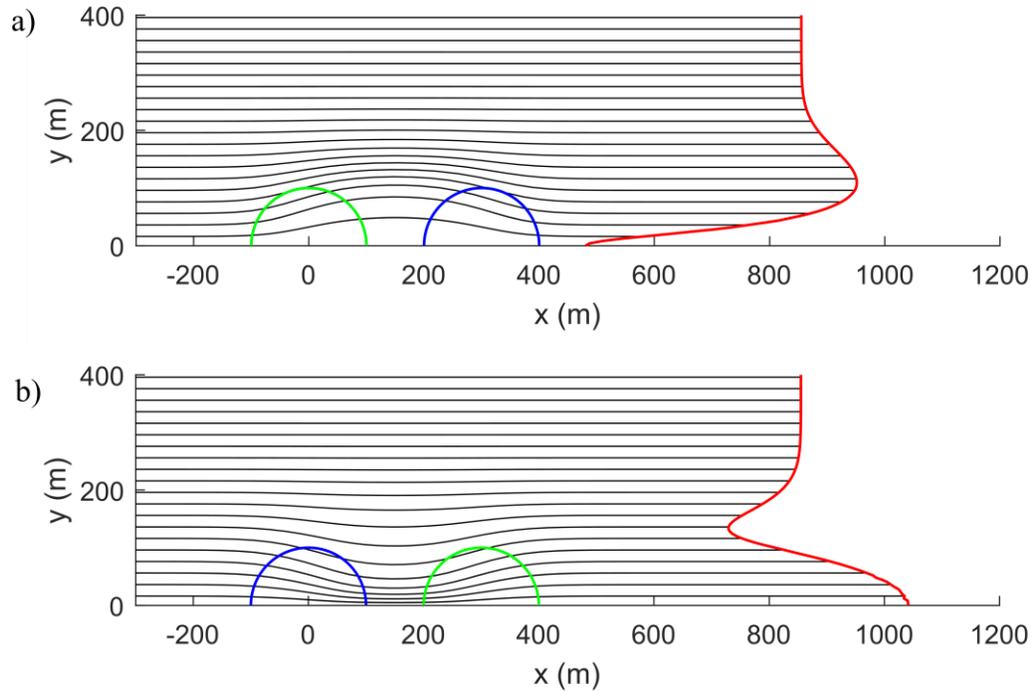

**Figure 2:** Characteristic solutions to (4) at $t = 14$ hours for **(a)** a Gaussian mound (green semicircle) at the origin followed by a Gaussian depression (blue semicircle) at $x = 300$ m, and **(b)** a Gaussian depression at the origin followed by a Gaussian mound at $x = 300$ m. Each have height $D = 6$ m and radius $L = 100$ m, and flow is from a line source at $x = -300$ m, where lava depth is 5 m. Note how the spacing of characteristics (and thus the flow depth and speed) returns to its unperturbed value after both orderings, but flow along $y = 0$ is advanced by (b) while it is delayed by (a). This is due to flow thinning between the features in (a).

The implication of this result is that, for modelling purposes, smoothing out a surface that contains many small mounds and depressions close together does not affect the steady flow on length scales much greater than the spacing of the small features. Hence, the study of flows over idealised topographic features is applicable to describing natural lava flows over roughened topography with the same long-wavelength form. Rumpf et al. (2018) showed experimentally that lava flows travel slower over roughened surfaces, which corroborates our result that the shape of the flow front may be a better indicator of upslope topography than the flow depth.

### 3.2 Concentric Mounds and Depressions

Having considered mounds and depressions speparately and in series, we turn our attention to axisymmetric features that combine the two. We study the effect of a circular elevation containing a depression at its centre on a plane inclined at angle $\boldsymbol{\beta}$ to the horizontal, such that the topographic elevation takes the form

$$E(x, y) = -x \tan\beta + D\left[\frac{x^2+y^2}{L^2}\right]\exp\left[-\frac{x^2+y^2}{L^2}\right]. \tag{8}$$

The feature's aspect ratio, $D/L$, is varied relative to the gradient of the slope, $\tan\beta$ to explore the effect of changing the relative amplitude of the feature. To determine if and where the surface is uphill relative to gravity, corresponding to $\partial E/\partial x > 0$, we consider different values of the dimensionless parameter

$$M = \frac{D}{L\tan\beta}. \tag{9}$$

This parameter quantifies the gradient of the feature relative to the underlying slope. The behaviour of the flow falls into one of three regimes, depending on the dimensionless parameter $M$. Although we generally use $\beta = 10°$, the qualitative results depend on $\beta$ only through the parameter $M$. The locations at which the surface is first uphill as $M$ is increased are at the steepest uphill points of the function $g(x) = x^2 \exp(-x^2)$, which are, to four significant figures,

$$x_1 = \left[\frac{5-\sqrt{17}}{4}\right]^{\frac{1}{2}} = 0.4682 \quad\text{and}\quad x_2 = -\left[\frac{5+\sqrt{17}}{4}\right]^{\frac{1}{2}} = -1.510. \tag{10}$$

For small values of $M$, the surface is nowhere uphill (as in figure 3a). As $M$ is increased beyond $M_1 = 1.703$ the surface is uphill near $x_1$ in the central depression (as in figure 4a). As $M$ is further increased beyond $M_2 = 2.529$, the surface also becomes uphill near $x_2$ on the feature's elevated margins (as in figure 5a). These critical values give rise to three regimes, described in the following subsections, for each of which the flow behaviour is qualitatively different.

### 3.2.1 Relatively Small Feature, $M < M_1$

In the low-amplitude regime, the topographic surface is nowhere uphill (as indicated in figure 3a for $M = 0.5$) and the characteristic model can be applied. Figure 3b shows that flow is slowed and diverted away from the elevated part of the feature, but accelerated and focussed in the central depression and at the margins of the feature. In the characteristic model, the flow depth and velocity are both permanently perturbed downstream of the feature, while in the full numerical solution these peturbations slowly decay downstream of the feature due to the smoothing effect of the second term in (2). Shallower upstream flow causes the downstream depth perturbation to decay over a longer length scale, indicating a scaling of the smoothing effect with $h_\infty$, as seen by comparison of the downstream length of the depth perturbations in figure 3c, where $h_\infty = 5$ m and figure 3d, where $h_\infty = 2.5$ m.

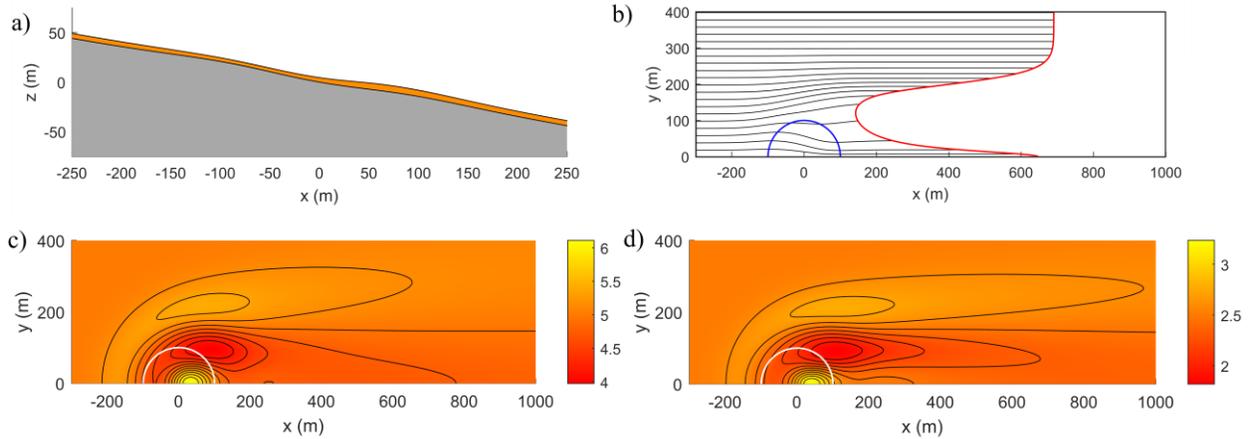

**Figure 3:** Flow over a low-amplitude feature described by (8) with $M = 0.5, L = 100$ m, $\beta = 10°$ and $h_\infty = 5$ m. **(a)** Cross-section of the topography and the fluid depth along $y = 0$ shows the feature is nowhere uphill. **(b)** Characteristic solution to (4) at $t = 12$ hours shows thinning and slowing at $y \sim L$ and thickening and acceleration at $y \sim 0$ and for $y > L$. Semicircles indicate feature radius. Numerical solutions to the steady state problem (2) for upstream flow depth **(c)** $h_\infty = 5$ m and **(d)** $h_\infty = 2.5$ m, where colour represents steady flow thickness in metres.

### 3.2.2 Ponding in the Central Depression $M_1 < M < M_2$

In the intermediate-amplitude regime, the topographic surface is uphill only in the central depression (as shown by the blue line in figure 4a for $M = 2$). The characteristic model fails there because characteristics converge to a point with 'infinite' thickness, as indicated in figure 4b. This breakdown indicates the formation of a deep pond of lava, with a horizontal free-surface. Although the flow will eventually overtop this feature, the timescale for overtopping must consider the time taken for the pond to fill. This depends on the volume of the filled pond, $V$, as well as the volumetric flux into the pond, which is determined by considering the width of the upstream domain from which characteristics converge in the pond (denoted by $-y_c < y < y_c$) and the flux per unit width, $Q$. The timescale for filling the pond and overtopping the feature, $t_{\text{fill}} = V/(2Qy_c)$, gets very long as $M \to M_2$ because as the upstream slope tends towards becoming uphill, more and more characteristics get diverted around the feature, such that $y_c \to 0$. Shown in figure 4d is the numerically determined relationship between $M$ and $t_{\text{fill}}$ and the least-squares fit for a curve of the form $t_{\text{fill}} = a(M_2 - M)^{-b}$. The fitting parameters were determined to be $a = (0.060 \pm 0.005)$ days and $b = 3.3 \pm 0.5$ with uncertainties indicating 95% confidence intervals, suggesting a roughly inverse cubic relationship between $t_{\text{fill}}$ and $(M_2 - M)$.

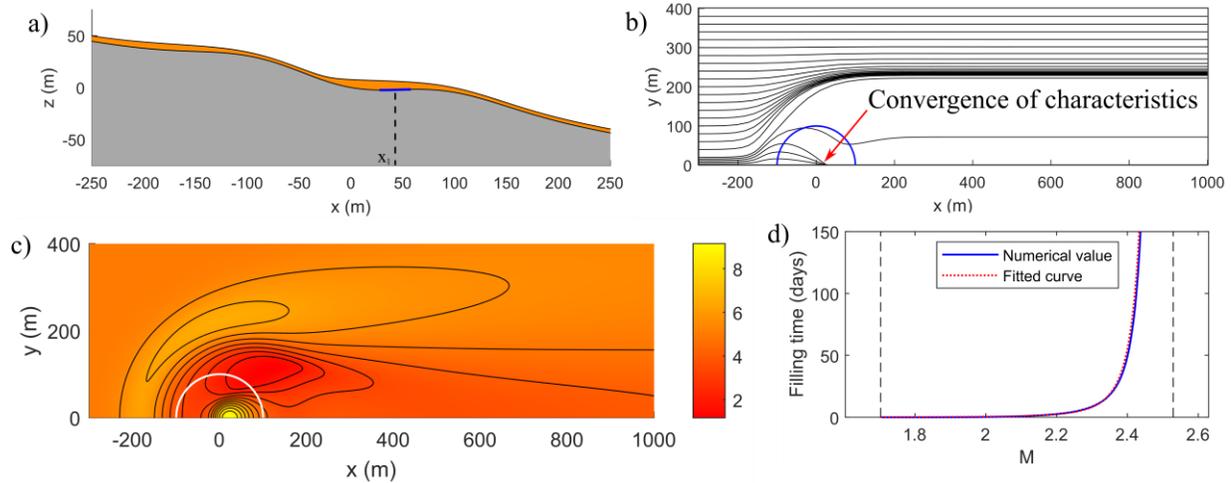

**Figure 4:** Flow over an intermediate-amplitude feature described in (8) with $M = 2, L = 100$ m, $\beta = 10°$ and $h_\infty = 5$ m. **(a)** A cross-section along $y = 0$ shows this feature is uphill only in the central depression (coloured blue), where a deep pond of lava with a horizontal free-surface forms. **(b)** Steady characteristic solution to (4) shows model breakdown by convergence of characteristics in the uphill region, indicating lava ponding. **(c)** Numerical solution to the steady problem (2), where colour represents steady flow thickness in metres. **(d)** The timescale for filling the pond, $t_{\text{fill}}$, as $M$ is increased from $M_1$ to $M_2$ (indicated by dashed vertical lines). The blue curve shows the numerically calculated values and the red dotted curve is a least-squares fit.

If the timescale for lava to pass over a feature is longer than the duration of an eruption, a region downstream of the feature may still be safe from inundation even if no lava-free region is predicted at steady state. In other words, topography need only be able to hold off the flow for the duration of an eruption, and not necessarily until steady state is reached. The ability of deep depressions to delay a flow has previously been exploited with limited success (Lockwood & Torgerson, 1980). The intermediate regime may be preferential behaviour for a barrier, because there is less flow deepening and acceleration at the margins of the feature than for the high-amplitude regime (discussed below). The lava that ponds within a depression may partially solidify during a flow, though this may have limited effect on $\boldsymbol{t_{fill}}$.

### 3.2.3. Lava-free Region, $M < M_2$

In the high-amplitude regime, the topographic surface is also uphill on the feature's elevated margins, as shown in figure 5a for $M = 3$. This leads to further breakdown of the characteristic model, because no characteristics access a large region containing much of the feature and its wake, as shown in figure 5b. This indicates a lava-free region, which extends downstream of the feature for a distance inversely proportional to the upstream flow depth, $\boldsymbol{h_\infty}$, as found by Hinton et al. (2019). The closure of the lava-free region is due to the second term in (2) which describes smoothing of gradients in $h$. The characteristic model ignores this smoothing effect, so does not predict the closure of the lava-free region. This behaviour is captured in the numerical solutions to (2), for different upstream flow depths, shown in figures 5c and 5d.

The timescale for overtopping this feature is now infinite, because no flow can make it over the feature, so infrastructure in its wake will be safe irrespective of the duration of the flow. The cost of this lava-free region, however, comes in the form of faster and deeper flow at its margins. It should be noted that ponded lava behind such an elevated feature may fully or partially solidify, especially in the common situation where lava supply is pulsatory. This modifies the shape of the topography such that future flows may overtop the feature more easily. This effect is beyond the scope of our present study.

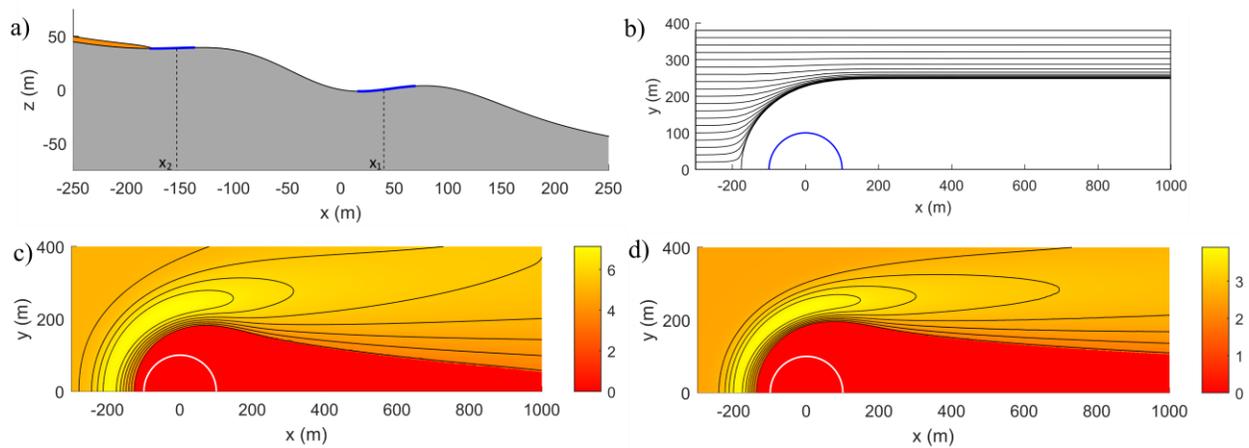

**Figure 5:** Flow over a high-amplitude feature described in (8) with $M = 3$, $L = 100$ m, $\beta = 10°$ and $h_\infty = 5$ m. **(a)** A cross-section along $y = 0$, showing this topography is uphill in the central depression and on the feature's elevated margins (both coloured blue). Lava is entirely diverted around this feature, as indicatd by the truncation of the orange layer. **(b)** Steady characteristic solution to (4) shows further model breakdown, because characteristics do not access a large region of the feature and its wake, indicating the formation of a lava-free region. Numerical solutions to the steady problem (2) for upstream flow depths **(c)** $h_\infty = 5$ m and (d) $h_\infty = 2.5$ m, where colour represents steady flow thickness in metres. These suggest that this lava-free region indeed closes downstream over a lengthscale inversely proportional to $h_\infty$.

### 3.3 Summary of Idealised Features

In this section, we have found that a depression and a mound in series cancel out each other's effects on steady flow thickness, but not on the speed of the flow front. We have also found that the flow front is accelerated downslope of depressions and at the margins of mounds, whilst it is slowed by mounds and at the margins of depressions. The acceleration effect at the margins of mounds is more pronounced by larger features, which has implications for lava hazard (as slowing flow in one location invariably leads to accelerating it in another). The effect of using mounds to slow down the flow front can be further increased by introducing deep depressions within mounds. Importantly, we suggest that investigating flow over an idealised topographic feature is relevant to the study of flow over realistic, roughened topography because small, closely spaced mounds and depressions are shown to cancel out perturbations in flow thickness.

## 4 Application to Real Topography

To verify the relevance of these conclusions to lava flows, the characteristic model and the full numerical model are applied to an a'ā flow 35 kyr ago at Marcath Volcano, Lunar Crater volcanic field, Nevada, where a lava-free region has been observed downslope of a topographic high. As visible in figure 6a, an elevated older volcanic cone caused the lava flow to split into northern and southern channels, which rejoin downslope of the older cone and feed several lobes which spread across the low-relief plane to the west. Although the eventual halting of the flow on the low-relief plane was likely cooling-limited, the behaviour of the flow around the older cone was largely controlled by the underlying topography.

The lava field is well preserved and has been measured using LiDAR by Younger et al. (2019), who recorded the extent of the erupted flows and made estimates of upper and lower bounds on the flow's effusion rate and bulk dynamic viscosity and an estimate of lava density, presented in table 2. Younger et al. (2019) also estimate the average lava thickness, $\bar{h} \sim 11$ m, which is similar to the lava heights measured at the margins. It is not clear whether the lava field at Marcath advanced in a single episode or as multiple flows. As a first approach, we use our model to simulate lava emplacement as a single episode, where the flow splits into multiple sheets to pass around the topographic high. Future work will update the model to simulate episodic effusion and solidification of lava between flows, which may be a more realistic representation of some a'ā flows (Wadge, 1978, Guest et al., 1987).

A digital elevation model (DEM) of a 3.2 km by 2.5 km rectangular region containing the lava field was obtained from the United States Geological Survey Shuttle Radar Topography Mission with one arc-second resolution (roughly 30 metres lateral resolution). This DEM describes the post-eruptive topography. In order to model the flow, pre-eruptive paleotopography is reconstructed by subtracting from the DEM an approximation of the flow thickness. For this, the flow thickness is set to 11m inside the outline of the flow and zero elsewhere, then smoothed using a Gaussian filter with radius 25 metres, following the observations of Younger et al. (2019). The palaeotopography is then smoothed with a Gaussian filter of radius $R$ metres. A contour map of the palaeotopography with $R = 25$ m is presented in figure 6b. The x- and y-axes represent N-S and E-W directions and the origin is at **115°59′30″** W 38°29'15″**N**. It should be noted that, for most volcanically active regions, higher resolution DEMs are available. The resolution of the model and the DEM must be exceeded by the wavelength of the feature that is to be reconstructed. At Marcath, we are interested in reproducing a lava-free region on the kilometer scale, so an arc-second resolution DEM is sufficient.

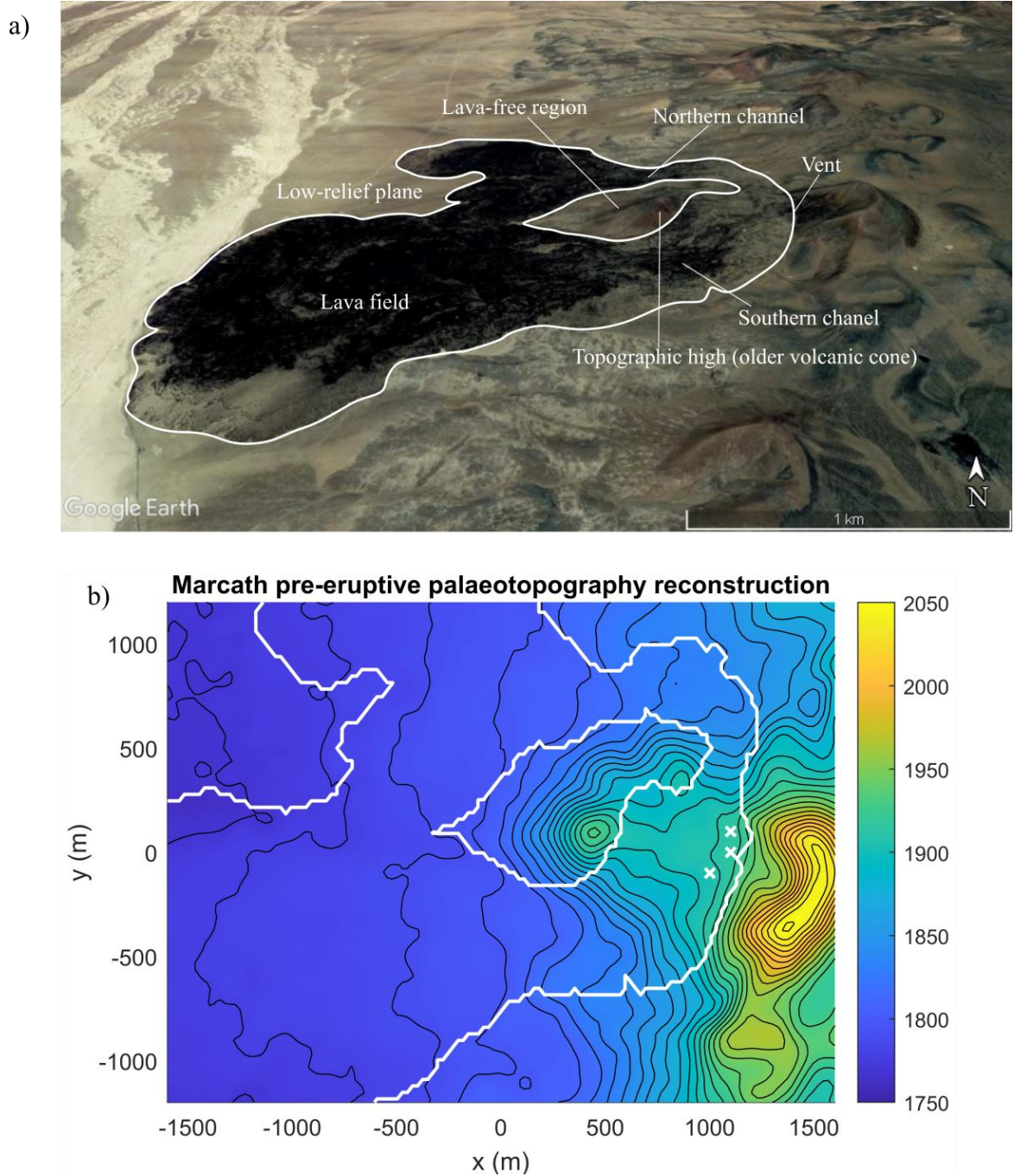

**Figure 6: (a)** Aerial view of Marcath volcano and lava field. The outline of the lava field is indicated by the white curve. There is a lava free region downslope of a topographic high. Map data: Google Earth, Copernicus, Landsat (2016). **(b)** Contour map of the reconstructed pre-eruptive palaeotopography, smoothed using a Gaussian filter with $R = 25$ m. Colour indicates elevation above sea level in metres and contour spacing is 10 metres. The observed lava field is outlined in white and the locations of the point sources used in the full numerical simulation are indicated by white crosses. Elevation data are obtained from USGS Shuttle Radar Topography Mission. Observational lava field data are taken from Younger et al. (2019).

For this topography, we employ the two methods used in Section 3. We solve equation (1) numerically and deploy the characteristic approximation (4) (see figure 7) over a time interval sufficiently long for the lava-free region to become enclosed (corresponding to ~35 km³ of lava extrusion, roughly half the total erupted volume). Details of the numerical method used are provided in the appendix. The vent from which the lava is extruded at Marcath is localised and cannot be modelled as an infinite line source as was done for the idealised topographies in section 3. For the purpose of the characteristic model, a 300 metre line source with lava depth $h_\infty = 11$ m is used, while in the full numerical model, lava is injected from three point sources at a total rate $q$ m³s⁻¹. The locations of these point and line sources are indicated in figure 6b. Approximations of the lava flux, viscosity and density made by Younger et al. (2019) are presented in table 2. The 'best' values are those used in our 'best' solution, while the uncertainties are used to probe the sensitivity of the model to parameters. The edge of the lava flow is selected to be the contour beyond which h begins to rapidly decrease. This is found to correspond to $0.7 \times$ (average thickness). The results are compared to the observed lava field shape in figure 7.

| Parameter | Quantity | Range | 'Best' value |
|---|---|---|---|
| $\mu$ | Dynamic viscosity of lava | $2 - 10$ MPa s | 5 MPa s |
| $q$ | Flux of lava from vent | $40 - 100$ m³s⁻¹ | 80 m³s⁻¹ |
| $\rho$ | Density of lava | 2,650 kg m⁻³ | 2,650 kg m⁻³ |

**Table 2:** An overview of the relevant model parameters from the most recent lava flow at Marcath; values reflect estimates taken from Younger et al. (2019). The viscosity estimates fall within the range suggested for a'ā lavas by Belousov and Belousova (2018).

### 4.1 'Best' simulation

The 'best' simulation was found by comparing the shape and thickness of the modelled and observed lava field shapes for different values of the parameters $\mu$ and $q$. The full numerical model has resolution 50 m, so topography is smoothed using $R = 25$ m, while the characterstic model uses $R = 100$ m. As noted in section 2, the shape of the lava field after the extrusion of a given volume of lava is determined only by $\lambda = q\mu/\rho$. The blending of $\mu$ and $q$ in this parameter means that there is an unresolvable uncertainty in estimates of the flux and viscosity for the eruption. This can only be resolved with multiple observations of the transient evolution to determine the flow timescale, which is unavailable for Marcath. Figures 7a-d show the evolution of the solutions to both the full numerical model and the characteristic approximation over the first 5 days of the eruption, with $q = 80$ m$^3$s$^{-1}$ and $\mu = 5$ MPa s. An animation of the simulation is provided in the supplementary information.

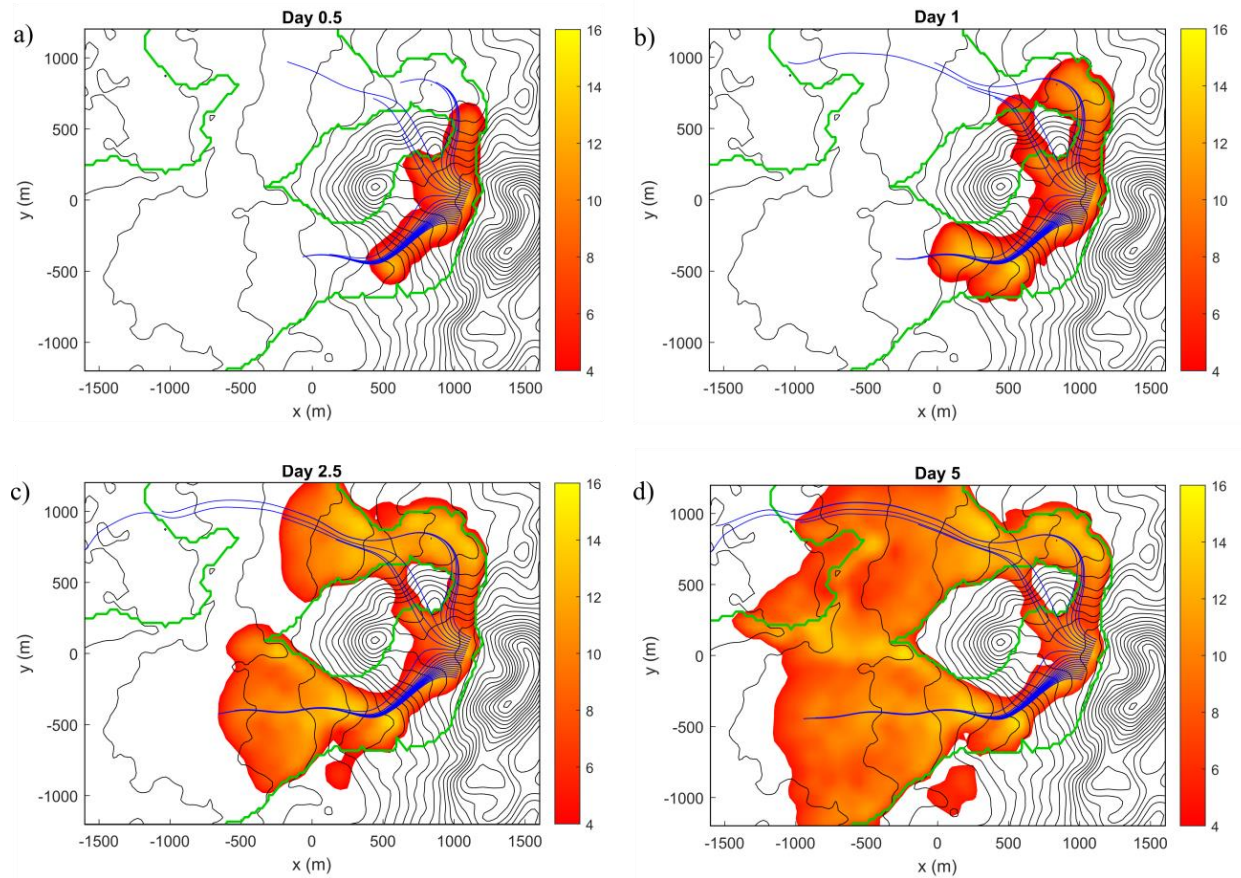

**Figure 7: (a-d)** Snapshots of the numerical solution to (1) taken at $t = 0.5, 1, 2.5$ and 5 days where the 'best' parameters are selected to be $q = 80$ m$^3$s$^{-1}$, $\mu = 5$ MPa s. Pre-eruptive palaeotopography is smoothed over $R = 25$ m and colour indicates lava thickness in metres. Overlain in blue are solutions to the characteristic model (4), with 20 characteristics emanating from a short line source at the vent. The observed lava-free region is represented by the green shape. The protrusion of the flow around $(0, -1000)$ m is an artefact of the selection of a cutoff $\boldsymbol{h}$, and is not a real feature of the lava flow.

Both the full numerical integration and the characteristic approximation successfully predict a lava-free region in the same location as the observed one. As shown in figure 7 and in the animation provided in the supplementary material, the simulations indicate that the lava initially fills the saddle between the vent and the topographic high to its west before splitting into branches and flowing around the topographic high. The branches rejoin on the downslope side of the elevation at roughly $t = 3.5$ days, after which the flow progresses along the low-relief plane to the west for the remainder of the simulation. The shape of the lava-free region is accurately reconstructed by the full numerical model, with the notable exception being the central branch of the lava flow. Field observations of the solidified lava field indicate that the flow split into two branches, while our simulation predicts three branches, one of which cuts the observed lava-free region. This discrepancy may be due to short-wavelength topography in the saddle between the vent and the topographic high that is not captured by the SRTM elevation data.

The characteristic model also predicts a lava-free region. Characteristics are focussed into three branches, much like the full numerical model. Although some characteristics appear to greatly outpace the rate of advance of the numerical flow front, the majority travel at a similar rate to the numerical solution. The main drawback of using this characteristic model to reconstruct lava-free regions is that the characteristics converge in valleys and stop entirely at local elevation minima. As is evident in figure 7, this means the shape of the lava flow in the far-field is poorly constrained by characteristics, as the branches of the flow do not all rejoin. The method of characteristics has some limited value in rapidly predicting flow behaviour downstream of topographic features.

### 4.2 Parameter sensitivity

To assess the performance of the full numerical model, its sensitivity to topographic smoothness and lava flux is investigated.

#### 4.2.1 Sensitivity to topographic smoothness

As found in section 3.1, smoothing topographic features over wavelengths shorter than the length scales of flow features we are trying to reconstruct should have little effect on steady flow depth. To investigate the effect of topographic smoothnes on our model, the reconstructed pre-eruptive palaeotopography is smoothed using a Gaussian filter with $R = 25 \, \text{m}, 50 \, \text{m}, 75 \, \text{m}$, and $100 \, \text{m}$. This amounts to smoothing over a sequentially longer wavelength.

The results of this are presented in figure 8. It is apparent that topographic smoothness indeed has little effect on average flow depth, nor on the shape of the flow front. Smoothing the underlying topography over wavelength $R$ largely acts to smooth the lava thickness over a similar wavelength. The shape of the lava-free region is reconstructed relatively well for all values of $R$ up to 100 metres. This is expected bcause the size of the lava-free region greatly exceeds 100 metres.

#### 4.2.2 Sensitivity to lava flux

Next, the sensitivity of the model to lava flux is investigated. The shape of the lava field after the effusion of a given volume of lava depends only on the topography and the parameter $\lambda = q\mu/\rho$, meaning increasing $q$ has an identical effect to increasing $\mu$. Hence this discussion also applies to changes in bulk lava viscosity. Younger et al. (2019) constrain the flux at Marcath to $40 - 100 \, \text{m}^3\text{s}^{-1}$. Here, we simulate flow with $q = 40 \, \text{m}^3\text{s}^{-1}, 60 \, \text{m}^3\text{s}^{-1}, 80 \, \text{m}^3\text{s}^{-1}$, and $100 \, \text{m}^3\text{s}^{-1}$ and the viscosity fixed at $\mu = 5 \, \text{MPa s}$. These simulations involve effusion of equal volumes of lava, so correspond to different times after the start of the eruption.

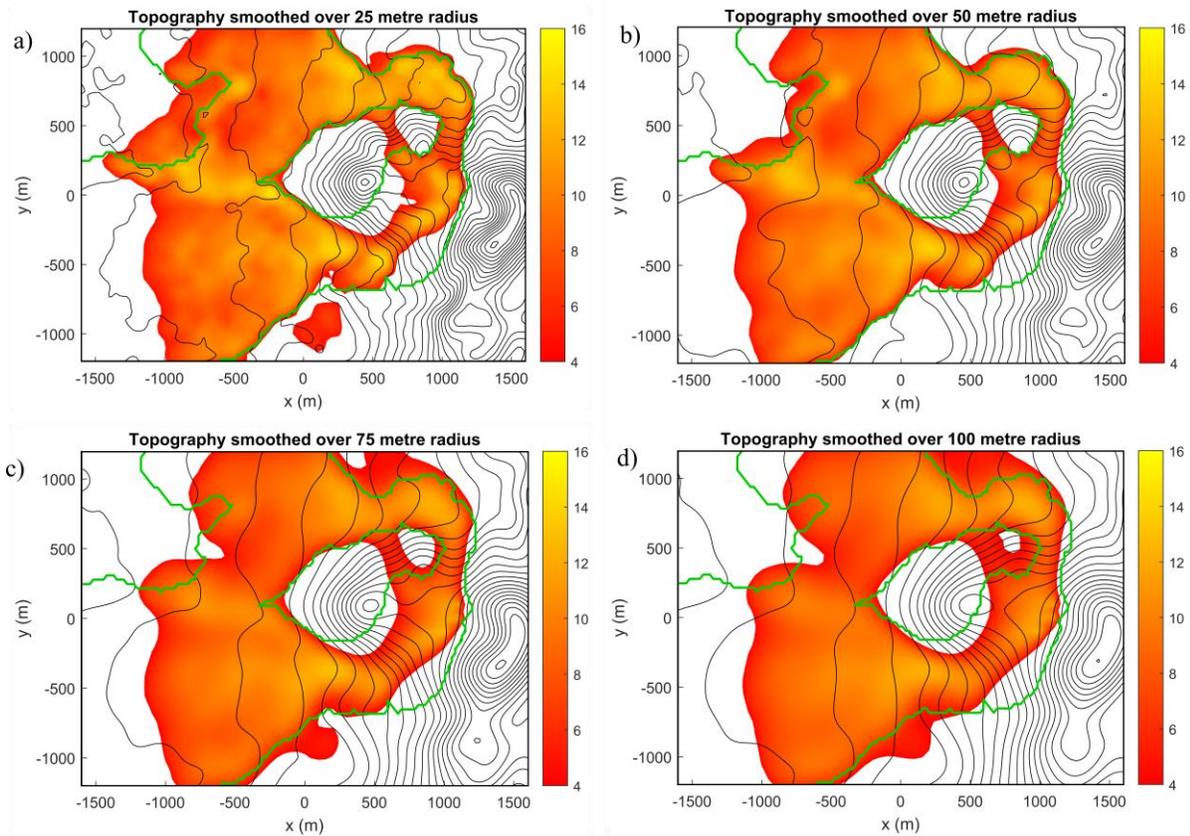

**Figure 8:** Numerical solutions to (1) for the Marcath eruption after effusion of 35 km³ of lava where palaeotopography is smoothed over radius $R$. **(a)** $R = 25$ m, **(b)** $R = 50$ m, **(c)** $R = 75$ m and **(d)** $R = 100$ m. Other parameters are fixed at $q = 80$ m³ s$^{-1}$, $\mu = 5$ MPa s, $\rho = 2650$ kg m$^{-3}$. Colour represents lava thickness in metres, the green shape represents the observed lava field.

The results of these evaluations are presented in figure 9. Within the range of flux studied, it is apparent that increasing the lava flux increases the average lava thickness without significantly altering the shape of the flow outline. A flow with greater flux will be thicker and will have inundated a slightly smaller area after effusion of a given volume of lava. The shape of the lava-free region is well reproduced for all investigated values of $q$.

The remarkable insensitivity of the lava-free region shape to topographic smoothness, lava flux and lava viscosity and the good match of the predicted lava-free region to field observations suggests that this method quickly and robustly captures the shape of lava-free regions, given topographic data. Only rough estimations of lava flux and viscosity are required, meaning this simple, quick method may be used for eruption forecasting, where such quantities are unknown.

On the other hand, the trade-off between flux and viscosity means there is uncertainty in the duration of the eruption owing to the dependence on the parameter $\lambda = q\mu/\rho$.

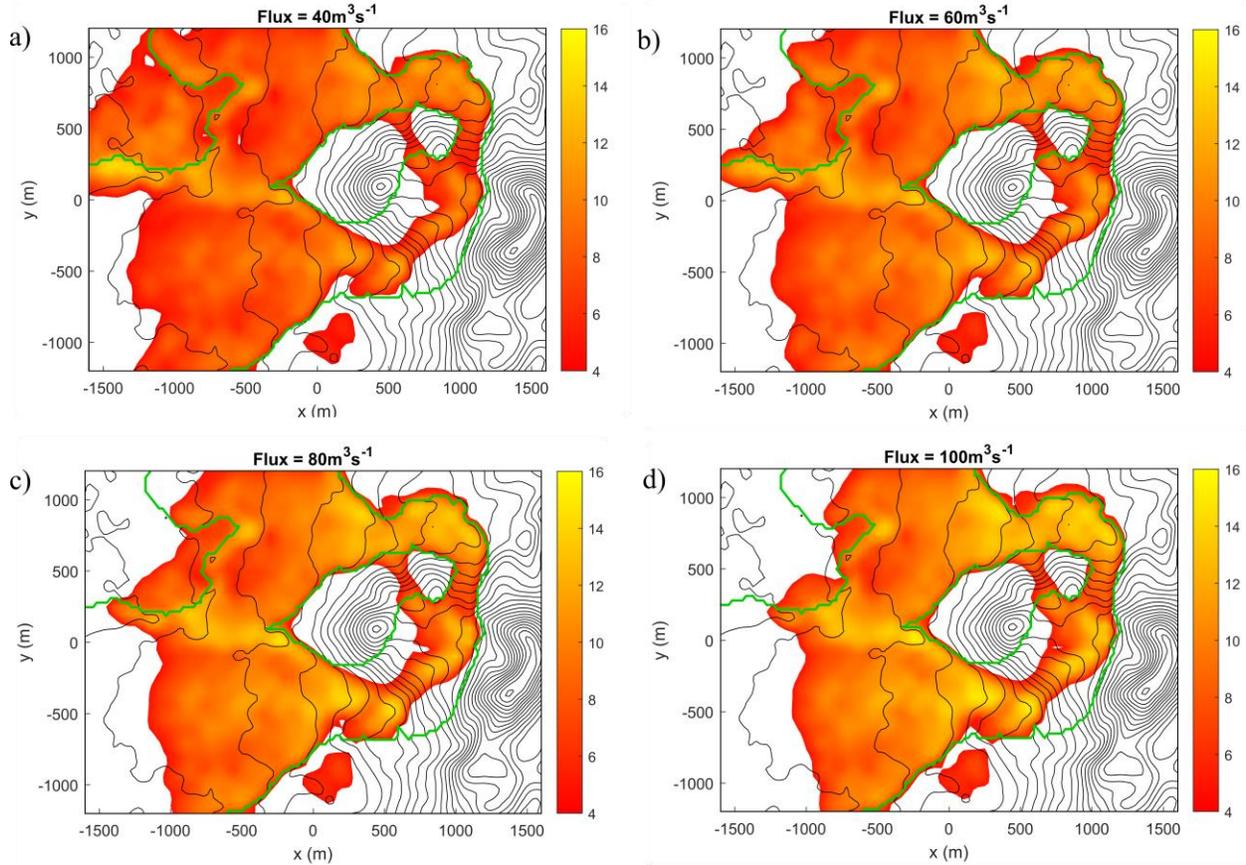

**Figure 9:** Numerical solutions to (1) for the Marcath eruption after effusion of 35 km$^3$ of lava, where dffusive flux, $q$, is varied. **(a)** $q = 40$ m$^3$s$^{-1}$, **(b)** $q = 60$ m$^3$s$^{-1}$, **(c)** q= 80 m$^3$s$^{-1}$ and **(d)** $q = 100$ m$^3$s$^{-1}$. Other parameters are fixed at $R = 25$ m, $\mu = 5$ MPa s, $\rho = 2650$ kg m$^{-3}$. Colour represents lava thickness in metres, the green shape represents the observed lava field.

## 5 Conclusions

In this paper, we have presented a model for the interaction of isothermal lava flows with general topography. Our approach is physically-motivated but sufficiently reduced to enable rapid computations. This efficiency is a major advantage for (i) timely forecasting of live erruptions, (ii) identification of the general flow physics and its dependency on the topography and input parameters and (iii) future work to predict key parameters from an observed flow (the inverse problem).

We have applied the model to idealized topographic features. We have found that topography is best characterized by the ratio of its local gradient to the gradient of the background slope. This has identified that features above a critical aspect ratio may not be overtopped by thin flows. Mounds divert the flow leading to thickened and accelerated motion at the margins, whilst depressions focus the flow. Identical mounds and depressions in series have a large influence on the transient flow front but little effect on the lava thickness further downstream. Hence, our work demonstrates that it is appropriate to use a mathematically idealised surface to investigate the effect of real topography, because multiple closely spaced small undulations on a surface have effects on steady flow depth that cancel out. This means that, at lengthscales longer than the wavelength of the undulations, a smoothed surface gives similar predictions for lava flow thickness as a rough one. We have determined that the effect of lava viscosity, $\mu$, density, $\rho$, and effusive flux, $q$, on the flow arises primarily through the parameter $\lambda = q\mu/\rho$. Furthermore, many of our results concerning the effect of topography apply to non-Newtonian lava.

We have also demonstrated how a deep depression within a small mound can slow the progress of the flow front without accelerating the flow at either side of the mound to the extent of a larger mound. This can give occupants more time to evacuate or protect the region behind the feature from the flow if the eruption duration is sufficiently short. Furthermore, we have shown that the transient flow front travels fastest at the margins of a mound. These places will be inundated by lava first and should be evacuated soonest in the case of a lava flow. We finally suggest that lava flow models should consider the effect of flow smoothing due to the component of gravity perpendicular to the surface, as well as downhill flow due to the component of gravity parallel to the surface. Doing so provides better constraint on the shapes of lava-free regions than the steepest-descent method allows.

The relevance of our model and each of these results has been corroborated by application to a lava flow at Marcath Volcano, Nevada, for which an animation of the modelled flow is available in the supplementary information. We suggest that our results be used to inform the safest positioning of key infrastructure with respect to natural topographic features: on or downslope of elevated topography. Incorporating the present analysis into volcanic hazard assessment models may improve the accuracy of prediction of the locations and sizes of safe zones, allowing homes and infrastructure to be better protected from inundation.

Future work will extend this approach to non-isothermal flows and eruptions whose effusion rate is not constant. This will allow more accurate modelling of real lava flows. Additionally, development of the ideas presented here could advise the design of artificial lava-diversion features. However, real lava ponded behind such a feature may cool and solidify enough that a subsequent flow may overtop the feature. Cooling and solidification within depressions is another promising avenue for research. Further future work could include inverting this model for an observed lava flow, to fit parameters such as vent location, effusion rate, lava viscosity and the form of the underlying topography.

## Data and Software

The elevation data used to produce the topography used in section 4 are obtained from the United States Geological Survey's Shuttle Radar Topography Mission 1 arc-second global dataset, entity IDs *SRTM1N37W117V3* and *SRTM1N37W116V3*. This is available at https://www.usgs.gov/centers/eros/science/usgs-eros-archive-digital-elevation-shuttle-radar-topography-mission-srtm-1 (doi:/10.5066/F7PR7TFT). The data describing the shape of the observed lava field at Marcath volcano (used in figures 6, 7, 8 and 9) is obtained from Younger et al. (2019).

The code used to produce the numerical results for steady and transient flows in this paper is available at https://github.com/Jacksaville1/lava-lubrication.

## Acknowledgements

The authors are grateful to E. Lev, G.B.M. Pedersen and R.S.J. Sparks for their insightful comments on this paper. J.M.S. would also like to thank the King's College Access Summer Internship fund for their generous funding, as well as colleagues A. Cox, M. Liu, M. Loncar, M. Roach, H. Rowland and O. Wilson for stimulating discussions and advice. E.M.H. is grateful to the University of Melbourne for the award of a Harcourt-Doig research fellowship.

## Appendix: Numerical Methods

In this appendix, we describe the numerical methods used in the paper for steady and transient lava flows.

The steady governing equation (2) is solved on a rectangular domain using MATLAB's Partial Differential Equation toolbox following the approach of Hinton et al. (2019). The boundary conditions are $h = h_\infty$ at the left boundary and $\partial h / \partial n = 0$ on the other three boundaries. The program uses finite elements to compute the solution. We use the initial guess that $h = h_\infty$ everywhere and iterate to obtain the flow thickness that accounts for the topography (see for example figure 3). In the case that there is a lava-free region, the method is adapted with a small source introduced to coat the lava-free region; for further details see section 3 of Hinton et al. (2019).

To solve the transient governing equation (1) numerically, the method of lines is used (Scheisser, 2012). First, the spatial derivatives are discretised using central differences. To accurately capture the steep regions of the free surface that arise at the edge of the flow, the KT scheme is applied to the hyperbolic term in the governing equation, $\nabla \cdot (h^3 \nabla E)$ (see Kurganov and Tadmor 2000). This discretization exploits a minmod flux limiter in the steep regions and limits the numerical viscosity. Details of the implementation can also be found in Appendix A of Zheng et al. (2015).This furnishes a system of ordinary differential equations for the evolution of the flow thickness at the spatial points. The system is stepped forward in time with the fourth-order Runge-Kutta method. It should be noted that this method handles lava-free regions effectively and hence no additional small sources are required. Checks were implemented to confirm that fluid volume is conserved. In addition, we found that the late-time solution of this transient solver converges to the steady solution of MATLAB's toolbox for the setup considered in figure 3, which acts as a check on our calculations